# Chandra observation of an ultraluminous X-ray source from the Galaxy NGC 1291

Juan C. Luna. [1]

*Department of Physics and Astronomy*

George Mason University, 4400 University Drive, Fairfax, Virginia 22030

## ABSTRACT

I report the analysis of an ultraluminous X-ray source (ULX) located in the galaxy NGC 1291. This X-ray point source is denominated IXO6 in the Catalog of Candidate IXO (Colbert & Ptak). An Intermediate-luminosity X-ray Object (IXO) is defined as an off-nuclear, compact object with luminosity Lx [2-10keV] $>= 10^{39}$ erg s-1. The cutoff Lx is defined as a value greater than the Eddington luminosity of a 1.4 M☉ black hole ($10^{38.3}$ erg s-1). IXO is an early denomination of what is call now a ULX point source. The Catalog was derived from a ROSAT survey and represents 87 IXOS in 54 galaxies. IXO6 was selected because of being positioned in the outer disk of the galaxy, with no near X-ray source neighbors. The study of this ULX pretends to confirm certain assumptions related to this class of objects (Roberts et al.)

## 1. INTRODUCTION

Observations of nearby galaxies with the Einstein X-ray satellite in the eighties showed X-ray sources with luminosities greater than black hole X-ray binaries (BH XRB) and less than active galactic nuclei (AGN).

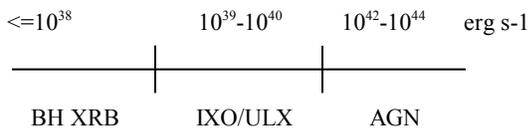

**Figure 1** Range of Energies

The nature of these sources is still unclear. Several hypotheses have been created to explain their range of luminosity. One of these hypotheses suggest that these objects are powered by accretion without beaming, therefore they would represent a new class of $10^2$-$10^5$ M☉ black holes, named intermediate-mass black holes (IMBH: Colbert & Mushotzky). The other hypothesis would be that this kind of X-ray luminosity is due to beamed emission from relativistic jets from high-mass binaries systems (HMXBR). And finally Supernova remnants are possible candidates to explain this particular range of Luminosity. Another hypothesis would be that we are actually observing micro-quasars (low Luminosity AGNs). At least two populations of ULX are known, some group of ULX was detected in active star formation regions, while another group was detected in elliptical galaxies more associated with older population of stars.

All of these hypotheses have pros and cons. Research is still going on in this particular kind of objects.

## 2. DATA REDUCTION

The galaxy NGC 1291 is an optically brightest Sa Galaxy with a bolometric magnitude of 9.39, distance of 8.9 Mpc, column density 2.1E+20 cm-2, and z=0.0280.

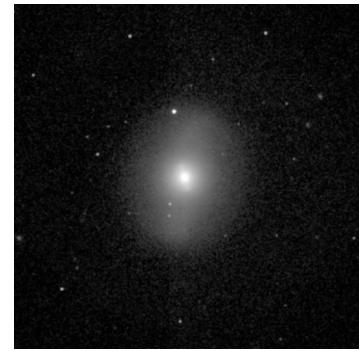

**Figure 2** Optical Image NGC 1291

The information of this galaxy was found in the NED (Nasa Extragalactic database) and using the FTOOLS utility 'nh', the Candidate Catalog was very helpful to locate the ULX candidate and to give some reference parameters to the data analysis.

I considered an observation made in epoch 2000 November 7, donated to the Chandra Archive by Andrea Prestwich (OBS. ID=2059). This observation has been taken during times of high background, in Data Mode Faint with the instrument ACIS-S. Our source IXO6 was present in the backside-illuminated S3 chip, and then most of the analysis was done on this particular chip.

---

[1]

[1] E-mail: jluna@gmu.edu

Level 1 files, and Level 2 files were processed to obtain the appropriate data products with CIAO 3.1. The original data was reprocessed with the latest version of CALDB (2.2.9), since the observation was using an old version of CALDB calibration (1.7).

Using the ACIS CCD numbering scheme the data was cleaned of flares and a lightcurve was created for every chip using the routine analyze_ltcrv.sl.

During the data processing WAVDETECT was used on reprocessed Level 2 data files to locate sources, and on-screen inspection was used to generate a clean file. A source file was created using DS9 to include or exclude regions near our point source.

A blank-sky file was created along with a spectral weights file using subprograms Sherpa, Chips and related S-lang routines.

The final result was a Xspec compatible file.

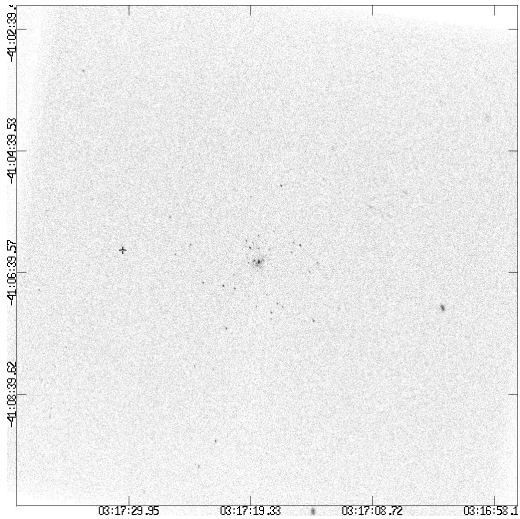

**Figure 3** Level 2 image NGC 1291

## 3. Spectral Analysis

The spectrum was analyzed in the 0.3-10keV energy range. A plot of the data without any modeling reveals a high component of soft and hard x-rays. The soft rays were supposed to be mainly a product of the hot gas in the galaxy, but its counts (>>50 ) were much greater than the accepted value of 50 for a hot ISM.

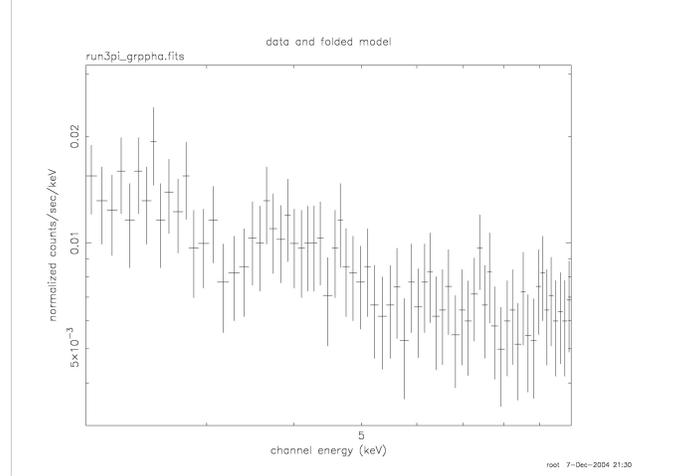

**Figure 4** IXO6 Spectra

The first model was a simple power law combined with an absorption model, the next model included a black body model, the one that was replaced by a bremstralung model. The fit of these models was very poor in the high energy region. The abundance of hard x-rays was explained by the pile-up effect (two or more photons detected as one photon). The next step was to fit a pile-up model with a power law spectrum.

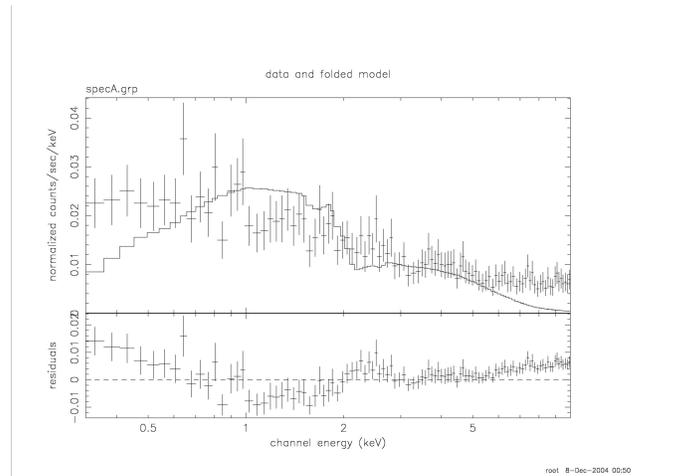

**Figure 5** Pile-up Fitted Model

The nature of the graph explains the difficulty of the fit with our current models. Related literature suggest fits with a DCBBM (disk color black body model) or Mekaal ( plasma model).

The luminosity calculated L= 0.3522 $10^{41}$ ergs/s is very well in the range of a ULX object, as well as the value of the flux: F= 2.3217E-04 photons ( 2.3707E-12 ergs)cm**-2 s**-1.

## 4. CONCLUSIONS

This point source is consistent with soft-spectrum sources like LMXBR, accretion powered BH or ULXs. I was able to confirm the high luminosity of this object as well as some specific features of its spectra, like shape, counts, flux, fitting models. I obtained a more accurate position of the object IXO6 (3:17:13,811, -41:07:27.23), this difference with the Rosat catalog was to be expected since Chandra has more spatial resolution. To advance in a further characterization of these objects, I propose a future catalog of ULX's spectra, that could consider time variability according with possible availability of observations in different epochs.